\documentclass[preprint,aps,floatfix,nofootinbib,a4paper]{revtex4-1}

\pdfoutput=1

\usepackage{amsmath, amssymb, amsfonts, amsthm, latexsym, epsfig, mathrsfs, xcolor, bbm, slashed}

\usepackage[all]{xy}

\usepackage[inline]{enumitem}

\usepackage{multirow}

\usepackage{setspace}
\usepackage[marginal, multiple]{footmisc}

\usepackage[T1]{fontenc}
\usepackage[utf8]{inputenc}
\usepackage{lmodern}

\usepackage[colorlinks, allcolors=blue!70!black, linktocpage]{hyperref}

\numberwithin{equation}{section}

\usepackage{cleveref}

\usepackage{microtype}

\usepackage{floatrow}

\let\OLDtableofcontents\tableofcontents
\renewcommand\tableofcontents[1]{%
    {\baselineskip 0.5ex %
	\OLDtableofcontents{#1}}%
}

\let\OLDthebibliography\thebibliography
\renewcommand\thebibliography[1]{%
	\setstretch{1.079} 
	\OLDthebibliography{#1}%
	\small %
	\setlength{\itemsep}{0.2\baselineskip} 
}

\let\OLDfootnote\footnote
\renewcommand\footnote[1]{%
	\setlength{\footnotesep}{0.75\baselineskip}%
	{\footnotesize \OLDfootnote{#1}}%
}

\setlist[enumerate]{noitemsep, label=(\arabic*), ref=(\arabic*)}

\newlist{condlist}{enumerate}{2}
\setlist[condlist,1]{noitemsep, topsep=0pt, label=(\arabic*), ref=(\arabic*)}
\setlist[condlist,2]{noitemsep, label=(\alph*), ref=(\arabic{condlisti}.\alph*)}
\crefname{condlisti}{condition}{conditions}
\crefname{condlistii}{condition}{conditions}

\newlist{propertylist}{enumerate}{1}
\setlist[propertylist,1]{noitemsep, topsep=0pt, label=(\arabic*), ref=(\arabic*)}
\crefname{propertylisti}{Property}{Properties}

\renewcommand\thesection{\arabic{section}}
\renewcommand\thesubsection{\arabic{subsection}}

\makeatletter
\def\p@subsection{\thesection.}
\def\p@subsubsection{\thesection.\thesubsection.}
\makeatother 

\theoremstyle{plain}

\theoremstyle{definition}

\theoremstyle{remark}

\crefname{section}{\S}{\S}
\crefname{appendix}{Appendix}{Appendices}
\crefname{figure}{Fig.}{Figs.}
\crefname{table}{Table}{Tables}

\crefname{definition}{Def.}{Defs.}
\crefname{prop}{Prop.}{Props.}
\crefname{lemma}{Lemma}{Lemmas}
\crefname{corollary}{Cor.}{Cors.}
\crefname{thm}{Theorem}{Theorems}
\crefname{remark}{Remark}{Remarks}

\crefname{ass}{Assumptions}{Assumptions}
\crefname{property}{Properties}{Properties}

\newcommand{\be}{\begin{equation}\begin{aligned}}
\newcommand{\ee}{\end{aligned}\end{equation}}

\newcommand{\lb}{\left}
\newcommand{\rb}{\right}

\newcommand{\ms}{\mathscr}
\newcommand{\mf}{\mathfrak}
\newcommand{\bb}{\mathbb}

\newcommand{\eqsp}{\, ,\quad} 

\newcommand{\hr}{\begin{center}* * *\end{center}}

\newcommand{\defn}{\mathrel{\mathop:}=} 

\let\oldsetminus\setminus
\renewcommand{\setminus}{\!\oldsetminus\!} 

\let\oldint\int
\renewcommand{\int}{\oldint\limits}

\let\oldlim\lim
\renewcommand{\lim}{\oldlim\limits}

\renewcommand{\bar}{\overline}

\newcommand{\scri}{\ms I}

\newcommand{\nfrac}[2]{{{}^#1\!\!/\!_#2}}

\newcommand{\half}{\nfrac{1}{2}}

\newcommand{\pb}[1]{\underleftarrow{#1}} 

\renewcommand{\Re}{{\rm Re\,}}
\renewcommand{\Im}{{\rm Im\,}}

\let\thorn\relax
\DeclareMathOperator{\thorn}{\text{\rm \th}}
\let\eth\relax
\DeclareMathOperator{\eth}{\text{\rm \dh}}

\newcommand{\wt}{\circeq}

\renewcommand{\o}{o}
\renewcommand{\i}{\iota}

\begin{document}

\setstretch{1.2}


\title{A twistorial description of BMS symmetries at null infinity}

\author{Kartik Prabhu}
\email{kartikprabhu@ucsb.edu}
\affiliation{Department of Physics, University of California, Santa Barbara, CA 93106, USA}

\begin{abstract}
We describe a novel twistorial construction of the asymptotic BMS symmetries at null infinity for asymptotically flat spacetimes. We define \emph{BMS twistors} as spinor solutions to some set of components of the usual spacetime twistor equation restricted to null infinity. The space of BMS twistors is infinite-dimensional. We show that given two BMS twistors their symmetric tensor product can be used to generate (complex) vector fields which are the infinitesimal BMS symmetries of null infinity. In this sense BMS twistors are ``square roots'' of BMS symmetries. We also show that these BMS twistor equations can be written a pair of covariant spinor-valued equations which are completely determined by the intrinsic universal structure of null infinity.
\end{abstract}

\maketitle
\tableofcontents

\section{Introduction}
\label{sec:intro}

For asymptotically flat spacetimes describing isolated systems in general relativity, it is well-known that at the asymptotic boundary, \emph{null infinity} denoted by \(\scri\), one obtains an infinite-dimensional asymptotic symmetry group --- the Bondi-Metzner-Sachs (BMS) group --- along with the corresponding charges and fluxes due to gravitational radiation \cite{BBM, Sachs1, Sachs2, Penrose, Geroch-asymp,GW, AS-symp, WZ}, see also \cite{GPS} for a recent exposition. On the other hand, the conformal geometry of flat Minkowski spacetime is elegantly described in terms of spinorial objects known as \emph{twistors} which satisfy a certain differential equation called the \emph{twistor equation} (\cref{eq:twis}), see \cite{PR2}.

However, it is well-known that there are significant difficulties in imposing the full twistor equation, even at null infinity, for general asymptotically flat spacetimes \cite{DS}. The usual approach to circumvent these issues, is to associate certain \emph{\(2\)-surface twistors} to a cross-section of null infinity \cite{Penrose-charges, DS, Shaw}. These \(2\)-surface twistors are required to satisfy only those components of the full twistor equation which are tangent to the chosen \(2\)-surface (\cref{eq:surface-twistor}). Using this approach, one can generate a vector field representing a Poincar\'e symmetry on some fixed cross-section of null infinity (see \cref{sec:twistors}). To this Poincar\'e symmetry one can associate a charge formula which is to represent the momentum and angular momentum of the spacetime at that cross-section \cite{Penrose-charges}. However, the restriction to a fixed cross-section and the resulting Poincar\'e symmetries is very unnatural from the point of view of asymptotic flatness and the universal geometric structure of null infinity.

In this work we obtain a new description of \emph{all} the asymptotic BMS symmetries starting from spinor solutions of a certain twistor-like equation at null infinity without choosing any particular cross-section (\cref{sec:bms-twistor}). Similar to the \(2\)-surface twistor approach we will only impose some components of the full twistor equation at null infinity (\cref{eq:bms-twistor}). Crucially, these will \emph{not} be all the components of the twistor equation which are tangent to \(\scri\). The solutions to these equations, which we call \emph{BMS twistors}, form an infinite-dimensional space. Using two such BMS twistors we then generate all BMS symmetries at null infinity (\cref{eq:twistor-to-vector}). The components of the twistor equations we use are precisely those that are determined only by the universal geometric structure of null infinity, and thus the BMS twistors also depend only on this universal structure. We make this precise in \cref{sec:tw-int} by writing these BMS twistor equations in an intrinsic covariant form (\cref{eq:twis-int}) which makes it manifest that they refer only to the universal structure at null infinity.

We conclude with a short discussion of some interesting new directions for future research suggested by this work in \cref{sec:disc}.

\section*{Notation and conventions}

We will use the definition of asymptotic flatness given by Penrose's conformal completion (see \cite{Geroch-asymp,Wald-book}), and denote null infinity as \(\scri \cong \bb R \times \bb S^2\). Abstract indices \(a,b,\ldots\) will be used for tensors in spacetime while \(A,B,\ldots\) and \(A',B',\ldots\) will be used for abstract spinor indices using the conventions in \cite{PR1}. We work exclusively in the conformally completed spacetime, the \emph{unphysical} spacetime \(M\) with a Lorentzian metric \(g_{ab}\). We use the mostly negative signature \((+,-,-,-)\) for the Lorentzian \(4\)-dimensional metric tensor \(g_{ab}\) on spacetime and denote the corresponding (antisymmetric) metrics on the spinor spaces by \(\epsilon_{AB}\) and \(\epsilon_{A'B'}\), see \cite{PR1}.

If \(\Omega\) is the conformal factor used to obtain the conformal-completion of the physical spacetime, then it can be shown that \(\nabla^a\Omega\) is the null generator of \(\scri\), and that one can, without loss of generality, choose \(\Omega\) so that the Bondi condition \(\nabla_a \nabla_b \Omega = 0\) is satisfied at \(\scri\). For intermediate computations, we will use the Geroch-Held-Penrose (GHP) formalism at null infinity \cite{GHP,PR1,PR2,KP-GR-match}. The GHP weight of any quantity \(\eta\) will be denoted by \(\eta \wt (p,q)\), and its spin will be \(s = (p-q)/2\). For this it will be convenient to make a choice of a null tetrad and spinor basis at \(\scri\) which determines a \emph{Bondi system}, see \cite{PR2} for details.

We pick a vector field \(n^a\) and a spinor \(\i^A\) at null infinity so that
\be\label{eq:normal-defn}
    A n^a = - \nabla^a\Omega \eqsp n^a = \i^A \i^{A'}
\ee
for some real function \(A\) with GHP weights \(A \wt (1,1)\).
Next, we pick a foliation of \(\scri\) so that the cross-sections are parallely-transported along \(n^a\). This foliation determines a unique null vector field \(l^a\) at \(\scri\) so that \(l_a = g_{ab}l^b\) is the conormal to the cross-sections and \(n_a l^a = 1\). Finally, we pick a complex null basis \(m^a\) and \(\bar m^a\) which is tangent to the cross-sections of this foliation and \(m_a \bar m^a = -1\). In this basis,
\be
    g_{ab} = 2 n_{(a} l_{b)} - 2 m_{(a} \bar m_{b)} \eqsp q_{ab} = - 2 m_{(a} \bar m_{b)}
\ee
where \(q_{ab}\) is the pullback of \(g_{ab}\) to \(\scri\), and is a (negative definite) Riemannian metric on the cross-sections of \(\scri\). We can also define another spinor \(\o^A\) so that \((\o^A, \i^A)\) and their complex conjugates \((\o^{A'}, \i^{A'})\) associated with the tetrads in the usual way (see \cite{PR2} for details) and normalized so that
\be
    \o_A \i^A = \o_{A'} \i^{A'} = 1
\ee
and all other contractions vanishing. In this choice of basis, the GHP spin coefficients at \(\scri\) satisfy
\be\label{eq:spin-scri}
    \kappa' &= \sigma' = \tau' = \rho' = \tau = \Im \rho = 0
\ee
while the spin coefficients \(\kappa, \sigma, \Re \rho\) are arbitrary. The function \(A\) appearing in \cref{eq:normal-defn} satisfies (see Eq.~9.8.26 of \cite{PR2})
\be\label{eq:A-relations}
    \thorn'A = \eth A = 0 \,.
\ee
Note that the spin coefficients \(\kappa\) and \(\Re \rho\) can also be set to zero, by appropriate choices of the conformal factor and tetrad away from \(\scri\), but we will not need to do so. The only nontrivial spin coefficient at \(\scri\) is \(\sigma\) which encodes the gravitational radiation through the News tensor, which is represented by a complex function \(N\) with
\be\label{eq:News-defn}
    \bar N \defn \thorn' \sigma
\ee

\section{Twistor equation at null infinity}
\label{sec:twistors}

In this section, we consider the twistor equation in the unphysical spacetime \(M\) evaluated at null infinity \(\scri\). A given spinor \(\omega^A\) satisfies the \emph{twistor equation} if
\be\label{eq:twis}
    \nabla_{A'}^{(A} \omega^{B)} = 0 \,.
\ee
The twistor equation is conformally-invariant: if we conformally rescale the metric \(g_{ab} \mapsto \varpi^2 g_{ab}\), where \(\varpi > 0\) is some smooth function, along with \(\omega^A \mapsto \omega^A\) then \(\nabla_{A'}^{(A} \omega^{B)} \mapsto \varpi^{-1} \nabla_{A'}^{(A} \omega^{B)}\) (see \S\ 6.1 of \cite{PR2}). Thus, solutions \(\omega^A\) of the twistor equation are conformally-invariant spinors on spacetime. For this reason, we will only need to consider the twistor equation on the unphysical spacetime \(M\) with some fixed choice of the conformal factor.

It will be convenient to decompose the spinor \(\omega^A\) into its components in a spinor basis as
\be\label{eq:omega-decomp}
    \omega^A = \omega^0 \o^A + \omega^1 \i^A
\ee
where the components have the GHP weights
\be
    \omega^0 \wt (-1,0) \eqsp \omega^1 \wt (1,0) \,.
\ee
We will frequently denote this decomposition as \(\omega^A \equiv (\omega^0, \omega^1)\). Note that \(\omega^0\) is spin \(s = -\half\) while \(\omega^1\) is spin \(s = \half\). The GHP form of the twistor equation can be found in Eq.~4.12.46 of \cite{PR1}, which when evaluated at \(\scri\) in our choice of frame gives
\begin{subequations}\label{eq:twistor-scri}\begin{align}
    (\thorn + \rho) \omega^0 = \eth'\omega^1 \eqsp \thorn \omega^1 = \kappa \omega^0 \label{eq:twis-ext} \\
    \eth \omega^1 = \sigma \omega^0 \label{eq:twis-rad} \\
    \eth' \omega^0 = 0 \eqsp \thorn' \omega^0 = 0 \eqsp \thorn' \omega^1 = \eth \omega^0 \label{eq:twis-univ} \,.
\end{align}\end{subequations}

\Cref{eq:twistor-scri} reveals that, in general, there are significant issues with imposing the twistor equation at \(\scri\), see also \cite{DS}. One cannot even impose the components of the twistor equation which are tangent to null infinity (\cref{eq:twis-rad,eq:twis-univ}). To see this, take the \(\thorn'\) of \cref{eq:twis-rad} and use the last equation in \cref{eq:twis-univ} to get
\be
    \eth^2 \omega^0 =  - \bar N \omega^0
\ee
However, since \(\omega^0\) is spin \(s = - \half\), the first equation in \cref{eq:twis-univ} implies that \(\eth^2 \omega^0 = 0\), and thus, we have \(\bar N = 0\) or \(\omega^0 = 0\). Clearly, the condition \(\bar N = 0\) is very restrictive as it demands that our spacetime have no radiation at null infinity. If we choose instead \(\omega^0 = 0\), then the remaining equations, \(\thorn'\omega^1 = \eth\omega^1 = 0\), have a \(2\) complex dimensional space of solutions from which we would not be able to generate the infinite-dimensional BMS symmetries --- these restricted solutions can be used to generate the \(4\)-dimensional space of BMS translations (see \cref{sec:bms-twistor}).\\

\hr

To side step these issues, the standard approach in twistor literature is to pick a fixed cross-section \(S\) of \(\scri\) and on this cross-section impose only the ``angular'' components of the twistor equation (see \cite{Penrose-charges, DS, Shaw}). That is, we impose
\be\label{eq:surface-twistor}
    \eth' \omega^0 = 0 \eqsp \eth \omega^1 = \sigma \omega^0 \quad \text{on } S \,.
\ee
These solutions define a \emph{\(2\)-surface twistor} \(\omega^A \equiv (\omega^0,\omega^1)\) at \(S\). Then, given two \(2\)-surface twistors \(\omega^A \equiv (\omega^0, \omega^1)\) and \(\tilde\omega^A \equiv (\tilde\omega^0, \tilde\omega^1)\), both solutions of \cref{eq:surface-twistor}, one defines a vector field \(\xi^a\) at the chosen cross-section \(S\) by
\be\label{eq:twistor-to-vector-S}
    \xi^a \big\vert_S \defn (A \beta)n^a + X m^a  \eqsp  \beta \big\vert_S = - i (\omega^0 \tilde\omega^1 + \omega^1 \tilde\omega^0) \eqsp X \big\vert_S = -2i A \omega^0 \tilde\omega^0
\ee
Using \cref{eq:surface-twistor}, one finds the conditions
\be\label{eq:surface-vector-cond}
    \eth' X \big\vert_S = 0 \eqsp \eth^2(A\beta) \big\vert_S = \tfrac{1}{2} \sigma \eth X + \eth(\sigma X)
\ee
Note that \(X\) has GHP weight \((-1,1)\) thus is spin \(s = -1\). Then, from prop.~4.15.58 of \cite{PR1}, \(\eth'X \big\vert_S = 0\) implies that \(X\) is supported only on the  \(\ell=1\) spin-weighted harmonics. Thus, there is a \(3\) complex dimensional space of \(X\), which span the Lie algebra \(\mf{sl}(2, \bb C)\), which is isomorphic to the Lorentz algebra. The second condition in \cref{eq:surface-vector-cond} is slightly trickier to interpret, but expresses a restriction of the BMS algebra to a particular Poincar\'e subalgebra (see \cite{DS}). For this choice of subalgebra, the (complex) charge associated with the symmetry \(\xi^a\) at \(S\) was proposed by Penrose \cite{Penrose-charges}. It can be shown that this charge formula is equal to the Wald-Zoupas charge formula \cite{WZ} for the choice of subalgebra of symmetries made above; see appendix~C.3 of \cite{GPS} and also \cite{DS,Shaw}.

The Poincar\'e subalgebra chosen using \cref{eq:surface-vector-cond} is quite unnatural from the point of view of null infinity. This choice cannot be transported away from the chosen cross-section \(S\), and different choices of cross-section give different Poincar\'e algebras, in general. So given an asymptotically flat spacetime, there is no natural choice for this Poincar\'e symmetry. Further, the Poincar\'e subalgebra chosen using \cref{eq:surface-vector-cond} is not universal since the choice depends on the shear \(\sigma\) which depends on the particular physical spacetime under consideration.

\section{BMS twistors and symmetries}
\label{sec:bms-twistor}

Now, we show that there exists an alternative strategy to side step the issues with imposing the full twistor equation at \(\scri\), which also helps generate the full set of BMS symmetries.

Our approach is to only impose the subset \cref{eq:twis-univ} of the twistor equation on \emph{all} of \(\scri\), that is
\be\label{eq:bms-twistor}
    \eth' \omega^0 = 0 \eqsp \thorn' \omega^0 = 0 \eqsp \thorn' \omega^1 = \eth \omega^0
\ee

It is quite straightforward to show that there are infinitely-many solutions to \cref{eq:bms-twistor}. Choose \emph{any} cross-section \(S \cong \bb S^2\) of \(\scri\) on which we will specify the ``initial'' values of \((\omega^0, \omega^1)\). Since \(\omega^0\) is spin \(s = -\half\), the first equation, \(\eth'\omega^0 = 0\) has a \(2\) complex dimensional space of solutions on \(S\); this follows from prop.~4.15.58 and table~4.15.60 of \cite{PR1}. In terms of spin-weighted spherical harmonics, \(\omega^0\) satisfying \(\eth'\omega^0 = 0\) is a spin \(s = -\half\) and \(\ell = \half\) function on \(\bb S^2\). The ``initial'' value of \(\omega^1\) is unconstrained except for being a spin \(s  =\half\) function on \(S\). The last two equations in \cref{eq:bms-twistor} can then be used to propagate this ``initial'' value to all of \(\scri\) along its null generators. Thus, we have an infinite-dimensional space of solutions \(\omega^A \equiv (\omega^0, \omega^1)\) to \cref{eq:bms-twistor} which we will call \emph{BMS twistors}.

Now let \(\omega^A \equiv (\omega^0 ,\omega^1)\) and \(\tilde\omega^A \equiv (\tilde\omega^0, \tilde\omega^1)\) be any two BMS twistors. Then, we define a vector field \(\xi^a\) on \(\scri\) by
\be\label{eq:twistor-to-vector}
    \xi^a = (A \beta)n^a + X m^a  \eqsp \beta = - i (\omega^0 \tilde\omega^1 + \omega^1 \tilde\omega^0) \eqsp X = -2i A \omega^0 \tilde\omega^0 \,.
\ee
This is similar to \cref{eq:twistor-to-vector-S}, except now both the BMS twistors and the vector field are defined on all of \(\scri\) instead of on a fixed cross-section. Using \cref{eq:bms-twistor} by direct computation one obtains
\be\label{eq:bms-cond}
    \thorn' (A\beta) = \half \eth X \eqsp \thorn' X = 0 \eqsp \eth' X = 0
\ee
These are precisely the conditions which define a (complex) BMS vector field on \(\scri\) (see \cite{DS} or appendix~B of \cite{GPS}). A general (complex) BMS vector field can be obtained as the linear span of such vector fields while a real BMS vector field can be obtained by taking the real part.

As before solutions of \(\eth'X = 0\) span the Lorentz Lie algebra \(\mf{sl}(2, \bb C)\), and \(\thorn'X = 0\) tells us how to propagate them along null generators of \(\scri\). Consider a particular solution \(\tilde\omega^0 = 0\), so that \(\thorn' \tilde\omega^1 = 0\). In this case, \(X = 0\) and \(\thorn' (A\beta) = 0\). Since the space of solutions \(\tilde\omega^1\) is infinite-dimensional, the space of solutions for \(\beta\) is also infinite-dimensional but is constant along the null generators of \(\scri\); these represent the (complex) BMS supertranslations. Finally, consider the case where \(\tilde\omega^0 = 0\) and \(\eth \tilde\omega^1 = 0\). Note that this last condition is now preserved under \(\thorn'\) even in radiative spacetimes where \(\bar N \not= 0\). Then since \(\tilde\omega^1\) is spin \(s = \half\), \(\eth\tilde\omega^1 = 0\) implies that it is supported only on the \(\ell = \half\) harmonics which is a \(2\) complex-dimensional space. In this case, the corresponding BMS vector field is
\be
    \xi^a = (A\beta) n^a \eqsp \beta = -i \omega^0 \tilde\omega^1 \eqsp \eth^2 (A\beta) = 0
\ee
Note that since \(\omega^0\) is \(s = -\half\) and \(\ell = \half\) and \(\tilde\omega^1\) is \(s = \half\) and \(\ell = \half\), \(\beta\) is \(s = 0\) and \(\ell =0\) or \(\ell = 1\), which is precisely the solutions to \(\eth^2(A\beta) = 0\) and represent the \(4\) complex BMS translations.

The charge and flux formulae associated with the BMS symmetries can be obtained using the Wald-Zoupas procedure \cite{WZ} (see appendix~B of \cite{GPS} for the expressions in the GHP notation), which can be written in terms of the BMS twistors using \cref{eq:twistor-to-vector}. Note that one cannot generalize our construction to obtain a ``quasi-local'' charge formula since the BMS symmetries are only defined at null infinity, and any extension of these symmetries into the spacetime is highly non-unique and gauge-dependent.

\section{Intrinsic and universal form of the BMS twistor equations on \(\scri\)}
\label{sec:tw-int}

The BMS vector fields on \(\scri\) are generators of diffeomorphisms which preserve the intrinsic universal structure at \(\scri\) \cite{Geroch-asymp,AS-symp}. We show in this section, that the BMS twistor equations can also be expressed as intrinsic universal equations on \(\scri\). Note we will retain the function \(A \wt (1,1)\) introduced in \cref{eq:normal-defn} to keep track of the GHP weights in our choice of tetrad basis; if one is concerned only with tensorial expressions then \(A\) can be set to \(1\).

Let us recall the ``first-order'' structure of \(\scri\) consists of a vector field \(A n^a\) and a degenerate Riemannian metric \(q_{ab}\), such that \(An^a q_{ab} = 0\). This structure is universal, in the sense that \(n^a\) and \(q_{ab}\) are intrinsically defined on the manifold \(\scri\), and are common to all asymptotically flat spacetimes. Different asymptotically flat spacetimes are instead distinguished by the ``second order'' structure encoded in equivalence classes of derivative operators on \(\scri\); we recall the essential aspects below and refer to \cite{Geroch-asymp,AS-symp} for details.

Let \(v_a\) be a \(1\)-form on \(\scri\) and let \(\mu_a\) be \emph{any} extension of \(v_a\) into the spacetime \(M\), i.e. \(\mu_a\) is a \(1\)-form in \(M\) such that \(v_a = \pb{\mu_a}\), where \(\pb{}\) denotes the pullback to \(\scri\). Then, a derivative operator \(D_a\) on \(\scri\) is defined as (see pp.~46 of \cite{Geroch-asymp})
\be
    D_a v_b \defn \pb{\nabla_a \mu_b}
\ee
Note that \(D_a\) is well-defined since it is independent of the choice of extension \(\mu_a\) of \(v_a\) into the spacetime \(M\), i.e. replacing \(\mu_a\) with \(\mu_a + \nu A n_a + \Omega \lambda_a\) does not affect \(D_a v_b\) on \(\scri\) \cite{Geroch-asymp}. Intrinsically on \(\scri\) this derivative operator satisfies
\be
    D_a (An^b) = 0 \eqsp D_a q_{bc} = 0
\ee

Two derivative operators \(\hat D_a\) and \(D_a\) are equivalent (they represent different conformal completions of the \emph{same} physical spacetime) if \cite{AS-symp}
\be
    (\hat D_a - D_a)v_b = f q_{ab} (An^c) v_c = - (\hat \rho - \rho) q_{ab} (An^c) v_c 
\ee
for some function \(f\) and \emph{all} \(v_b\) on \(\scri\). In our tetrad basis this function is given by the difference of the spin coefficient \(\rho\) as indicated above. Let us denote by \(\{D\}_a\) the equivalence class of the derivative operator \(D_a\) under the above equivalence relation.

The difference of equivalence classes of derivatives is given by a tensor \(\gamma_{ab}\)
\be
    \lb( \{\hat D\}_a - \{D\}_a \rb) v_b = \gamma_{ab} (A n^c) v_c  \eqsp \gamma_{ab} (An^b) = 0 \eqsp q^{ab} \gamma_{ab} = 0
\ee
for \emph{all} \(v_b\). In our tetrad basis this is
\be
    \gamma_{ab} = (\hat\sigma - \sigma) \bar m_a \bar m_b + \text{c.c.}
\ee
where \(\text{c.c.}\) denotes the complex conjugate of the previous expression. The shear spin coefficient \(\sigma\) encodes the different equivalence classes of derivatives and thus the radiative degrees of freedom at \(\scri\) \cite{AS-symp}.

Since the BMS twistor equations \cref{eq:bms-twistor} do not depend on the spin coefficients \(\rho\) and \(\sigma\), we can already see that the BMS twistors do not depend on the choice of derivative operator on \(\scri\). Thus, the BMS twistors are universally defined on the manifold \(\scri\), and are common to all asymptotically flat spacetimes. Since the BMS twistor equations are universal, can we express them entirely in terms of the intrinsic derivative operators \(D_a\) on \(\scri\)? As we show next, this is indeed possible.

First we note that since \(n^a = \i^A \i^{A'}\), we can consider \(\i^A\) and its conjugate as part of the universal structure. Secondly, we can easily extend the derivative operator \(D_a\) to act on spinor fields on \(\scri\). Consider the ``Infeld-van der Waerden symbols'' \(\sigma^a_{AA'}\) in \(M\) which are implicitly used to convert between a tensor index and a pair of spinor indices \cite{PR1}. At \(\scri\), we can express them in our tetrad and spinor basis as
\be
    \sigma^a_{AA'} = n^a \o_A \o_{A'} - m^a \i_A \o_{A'} - \bar m^a \o_A \i_{A'} + l^a \i_A \i_{A'}
\ee
Note that \(\sigma^a_{AA'}\) is a covariant map from tensor products of spinor fields to vector fields. Then the BMS twistor equations \cref{eq:bms-twistor} can be expressed covariantly as
\be\begin{aligned}
    0 & = \i_B \pb{\sigma_{a A A'}} \nabla^{A'(A} \omega^{B)} = - l_a \lb( \thorn' \omega^0 \rb) + m_a \lb( \eth' \omega^0 \rb) - \tfrac{1}{2} \bar m_a \lb( \thorn' \omega^1 - \eth \omega^0 \rb)
\end{aligned}\ee
where, as before, \(\pb{}\) denotes the pullback of the tensor index to \(\scri\). Note that in the second expression we have written the covariant form in our choice of basis. This cannot be directly expressed in terms of the intrinsic derivative \(D_a\) due to the appearance of a term \(\nabla^{A'B} \omega^A\) where the \(A\) index is not on the spacetime derivative.

However, lets define
\be\label{eq:Gamma-defn}
    \sigma^a_A &\defn \sigma^a_{AA'} \i^{A'} \eqsp \sigma^a_{A'} &\defn \sigma^a_{AA'} \i^A \, .
\ee
Since these quantities are tangent to \(\scri\), we can consider them as spinor-valued vector fields intrinsically on \(\scri\). By direct computation, they satisfy the identities
\be
    \bar{\sigma^a_A} = \sigma^a_{A'} \eqsp \sigma^a_A \i^A = n^a \eqsp q_{ab} \sigma^a_A \sigma^b_B = 0 \eqsp q_{ab} \sigma^a_A \sigma^b_{B'} = \i_A \i_{B'}
\ee
and their conjugates. Now, we use \(\sigma^a_A\) and \(\sigma^a_{A'}\) to define the spinor-valued derivatives
\be
    D_A \defn \sigma^a_A D_a \eqsp D_{A'} \defn \sigma^a_{A'} D_a 
\ee
with \(D_{A'} = \bar{(D_A)}\) and \(\i^A D_A = \i^{A'} D_{A'} \). In terms of the spacetime derivative, these are given by \(D_A = \i^{A'} \nabla_{AA'}\) and \(D_{A'} = \i^A \nabla_{AA'}\).

If \(\hat D_a\) and \(D_a\) are equivalent derivative operators on \(\scri\) then for any spinor \(\mu_A\) we have
\be\label{eq:D-equiv-spin}
    (\hat D_A - D_A) \mu_B = 0 \eqsp (\hat D_{A'} - D_{A'}) \mu_B = (\hat\rho - \rho) \i_{A'} \i_B \i^C \mu_C
\ee
while, the difference of equivalence classes of derivative operators is given by
\be\label{eq:D-class-spin}
    (\{ \hat D \}_A - \{ D \}_A) \mu_B = (\hat\sigma - \sigma) \i_A \i_B \i^C \mu_C  \eqsp ( \{ \hat D \}_{A'} - \{ D \}_{A'}) \mu_B = 0
\ee
The corresponding action on primed spinors are obtained by taking the complex conjugate of the above equations.\\

The BMS twistor equations can be now expressed as a pair of spinor-valued equations
\begin{subequations}\label{eq:twis-int}\begin{align}
    0 & = \i_B D^{(A} \omega^{B)} =  - \o^A (\thorn' \omega^0 ) - \tfrac{1}{2} \i^A \lb[ \thorn' \omega^1 - \eth \omega^0 \rb] \label{eq:twis-int-A}\\
    0 & = \i_B D^{A'} \omega^B = - \o^{A'} (\thorn' \omega^0 ) + \i^{A'} (\eth' \omega^0 ) \label{eq:twis-int-A'}
\end{align}\end{subequations}
Note that these form of the equations are completely covariant and intrinsically defined on \(\scri\); we have expressed them in our choice of basis for convenience. Using \cref{eq:D-equiv-spin,eq:D-class-spin}, it is straightforward to check that both \cref{eq:twis-int-A,eq:twis-int-A'} are invariant under any change of derivative operators and their equivalence classes. Thus, \cref{eq:twis-int} is the BMS twistor equation written in an universal form, completely intrinsically on null infinity. The BMS vector field (\cref{eq:twistor-to-vector}) obtained from two BMS twistors \(\omega^A\) and \(\tilde\omega^A\) can also be expressed covariantly as
\be
    \xi^a = 2i A \sigma^a_A \i_B \omega^{(A} \tilde\omega^{B)} \,.
\ee

\hr

Note that if \(\omega^A\) satisfies the usual twistor equation \cref{eq:twis} then, using the fact that the spinor space is \(2\)-dimensional, one can infer the existence of another spinor \(\pi_{A'}\) such that (see \cite{PR2})
\be\label{eq:pi-defn-bulk}
    \nabla_{A'A}  \omega^B = - i \epsilon_A{}^B \pi_{A'} \eqsp \pi_{A'} = \tfrac{i}{2} \nabla_{A'A} \omega^A \, ,
\ee
where the second equation follows from the first by taking a trace over the \(A\) and \(B\) indices.

One can do something similar for the BMS twistor equations (\cref{eq:twis-int}) as follows. First, \cref{eq:twis-int-A'} implies that there exists a spinor \(\pi_{A'}\) such that (note that this \(\pi_{A'}\) is unrelated to the one in \cref{eq:pi-defn-bulk})
\be\label{eq:pi-defn}
    D_{A'} \omega^B &= -i \pi_{A'} \i^B
\ee
Then, \cref{eq:twis-int-A} implies that there exists a \(\lambda\) such that
\be\label{eq:lambda-defn}
    D_A \omega^B &= - \lambda \i_A \i^B - i \pi^{0'} \epsilon_A{}^B
\ee
where we have used the fact that \(\i^A D_A = \i^{A'} D_{A'}\) and, in our notation, \(\i^{A'} \pi_{A'} = \pi^{0'}\). Using our choice of spinor basis we find
\be\label{eq:pi-lambda-basis}\begin{aligned}
    \lambda & = \eth \omega^1 - \sigma \omega^0 \\
    \pi^{0'} & = \tfrac{i}{2}D_A \omega^A =  \tfrac{i}{2} \lb[ \thorn' \omega^1 + \eth \omega^0 \rb] = i \eth\omega^0 \\
    \pi_{A'} & = i ~ \o_{A'} (\thorn'\omega^1 ) - i~ \i_{A'} (\eth'\omega^1 -\rho \omega^0) = i ~ \o_{A'} (\eth\omega^0 ) - i~ \i_{A'} (\eth'\omega^1 -\rho \omega^0)
\end{aligned}\ee 

Using \cref{eq:D-equiv-spin,eq:D-class-spin} and \cref{eq:pi-defn,eq:lambda-defn}, or the basis expressions \cref{eq:pi-lambda-basis}, it can be checked that if \(\hat D_a\) and \(D_a\) are equivalent derivative operators then
\be
    \hat\lambda = \lambda \eqsp \hat\pi_{A'} = \pi_{A'} + i (\hat\rho - \rho) \i_{A'} \omega^0 \eqsp \hat \pi^{0'} = \pi^{0'}
\ee
and similarly if \(\{\hat D\}_a\) and \(\{D\}_a\) are different equivalence classes of derivatives we have
\be
    \hat \lambda = \lambda - (\hat\sigma - \sigma) \omega^0 \eqsp \hat\pi_{A'} = \pi_{A'}
\ee
Thus, while \(\lambda\) and \(\pi_{A'}\) are not universal, \(\pi^{0'} = \i^{A'} \pi_{A'}\) is universally defined on \(\scri\).

\section{Discussion}
\label{sec:disc}

The relationship of these BMS twistors at null infinity with other aspects of twistor theory would be interesting to explore. We mention a few possible future directions.

The universal geometric structure of null infinity is a \emph{conformal Carroll structure} on \(\scri \cong \bb R \times \bb S^2\), which is an ``ultra-relativistic'' limit (speed of light tends to zero) of conformal Lorentzian structures \cite{LL,Duval:2014uva,Duval:2014lpa,Hartong:2015xda,Ciambelli:2019lap,Figueroa-OFarrill:2019sex}. In this sense, an ``ultra-relativistic'' limit of Ward's mini-twistor space \cite{Ward}, could describe some universal twistorial structure of \(\scri\) and might shed more light on the BMS twistors defined in this paper. While the ``non-relativistic'' limit of twistor theory has been investigated \cite{Dunajski:2015lxa}, we are not aware of any such work on the ``ultra-relativistic'' limit.

Limits of the twistor equation to spatial infinity have also been considered previously by Shaw \cite{Shaw-spatial-inf1,Shaw-spatial-inf2}. In this context, it has been recently shown that, for suitably regular spacetimes, the asymptotic BMS symmetries at \emph{both} future and past null infinites can be matched onto each other through spatial infinity \cite{KP-GR-match,PS-Lorentz-match}. The twistorial aspects of this matching and the relation to BMS twistors described in this paper are certainly worth investigating.

Finally, we note that the construction of the BMS twistors uses the universal structure at null infinity. It would be interesting to see if a similar construction can be carried out at finite null surfaces in general relativity using the universal structure defined in \cite{CFP} to generate symmetries at finite null surfaces.

\section*{Acknowledgements}
This work is supported by NSF grant PHY-2107939. Some calculations in this paper used the computer algebra system \textsc{Mathematica} \cite{Mathematica}, in combination with the \textsc{xAct/xTensor} suite~\cite{xact,xact-spinors}.




\bibliographystyle{JHEP}
\bibliography{twistors-bms}
\end{document}